\documentclass[sigconf]{acmart}

\AtBeginDocument{%
  }

\setcopyright{acmlicensed}
\copyrightyear{2018}
\acmYear{2018}
\acmDOI{XXXXXXX.XXXXXXX}
\acmConference[Conference acronym 'XX]{Make sure to enter the correct
  conference title from your rights confirmation email}{June 03--05,
  2018}{Woodstock, NY}
\acmISBN{978-1-4503-XXXX-X/2018/06}

\usepackage{enumitem}
\usepackage{graphicx}
\usepackage{subcaption}
\usepackage{multirow}
\usepackage{array}

\usepackage{algorithm}
\usepackage{algpseudocode}
\usepackage{makecell}
\usepackage{bm}

\begin{document}

\title[TM20K: Teacher Retains Full Tokens, Student Merges Efficiently]{Teacher Retains Full Tokens, Student Merges Efficiently: TM20K for E-Commerce Sequence Modeling in Ad Recommendation}

%

\author{Xinchun Li$^*$, Duoru Zheng$^*$, Wenlin Zhao$^*$, Haoran Ding$^*$, Ziyi Zhou$^*$, Jingxuan Tan$^*$, Huizhi Yang$^*$, Yuchen Jiang, Zhe Chen, Yuchao Zheng, Linlan Chen, Dongjian Wang, Dongyue Wang, Xiaosong Li, Hongyue Mao, Yaocheng Tan$^\dagger$}
\thanks{$^*$ Co-first authors, equal contributions. \\ $^\dagger$ Corresponding author.}
\affiliation{
  \institution{ByteDance}
  \country{\{lixinchun.bu,zhengduoru,zhaowenlin,dinghaoran,zhouziyi.828,tanjingxuan,yanghuizhi,jiangyuchen.jyc,chenzhe.john, zhengyuchao.yc,chenlinlan,wangdongjian.msg,wangdongyue,lixiaosong.1,maohongyue,tanyaocheng\}@bytedance.com}
}

\renewcommand{\shortauthors}{Li et al.}

\begin{abstract}
  Benefiting from ultra-long behavior sequence modeling, existing recommender systems bring users a better experience via simultaneously considering their long-term and short-term interests. Nevertheless, extended sequence lengths introduce substantial burdens on training efficiency and serving throughput. Prior approaches typically utilize search-based or cluster-based compression on ultra-long sequences at the cost of fine-grained information, or rely on various lightweight target attention structures incapable of sufficient sequential feature extraction. In this paper, we balance the effectiveness and efficiency for ultra-long sequence modeling via {\it full transformer modeling} accompanied with {\it a two-stage knowledge distillation framework}. First, both teacher and student models take the full attention mechanism rather than pure target-sequence attention for effective sequence scaling. For student models, we propose several {\it simple yet well-motivated token merge approaches}, significantly compressing the sequence length while maintaining an acceptable performance. Then, a {\it one-time} teacher is heavily trained {\it with full sequence tokens}, further boosting the performance of student models via knowledge distillation. The proposed paradigm named \textbf{TM20K} has been successfully deployed in ByteDance's e-commerce advertising recommender system that extends {\it the e-commerce sequence length to 20K}, delivering substantial improvements in key business metrics (e.g., {\it ADSS$+1.036\%$}) while keeping the training and serving cost nearly the same as the online state-of-the-art model (e.g., {\it serving latency only $+5.6\%$}).
\end{abstract}




\maketitle

\section{Introduction}

\begin{figure}[htbp]
  \centering
  \includegraphics[width=\linewidth]{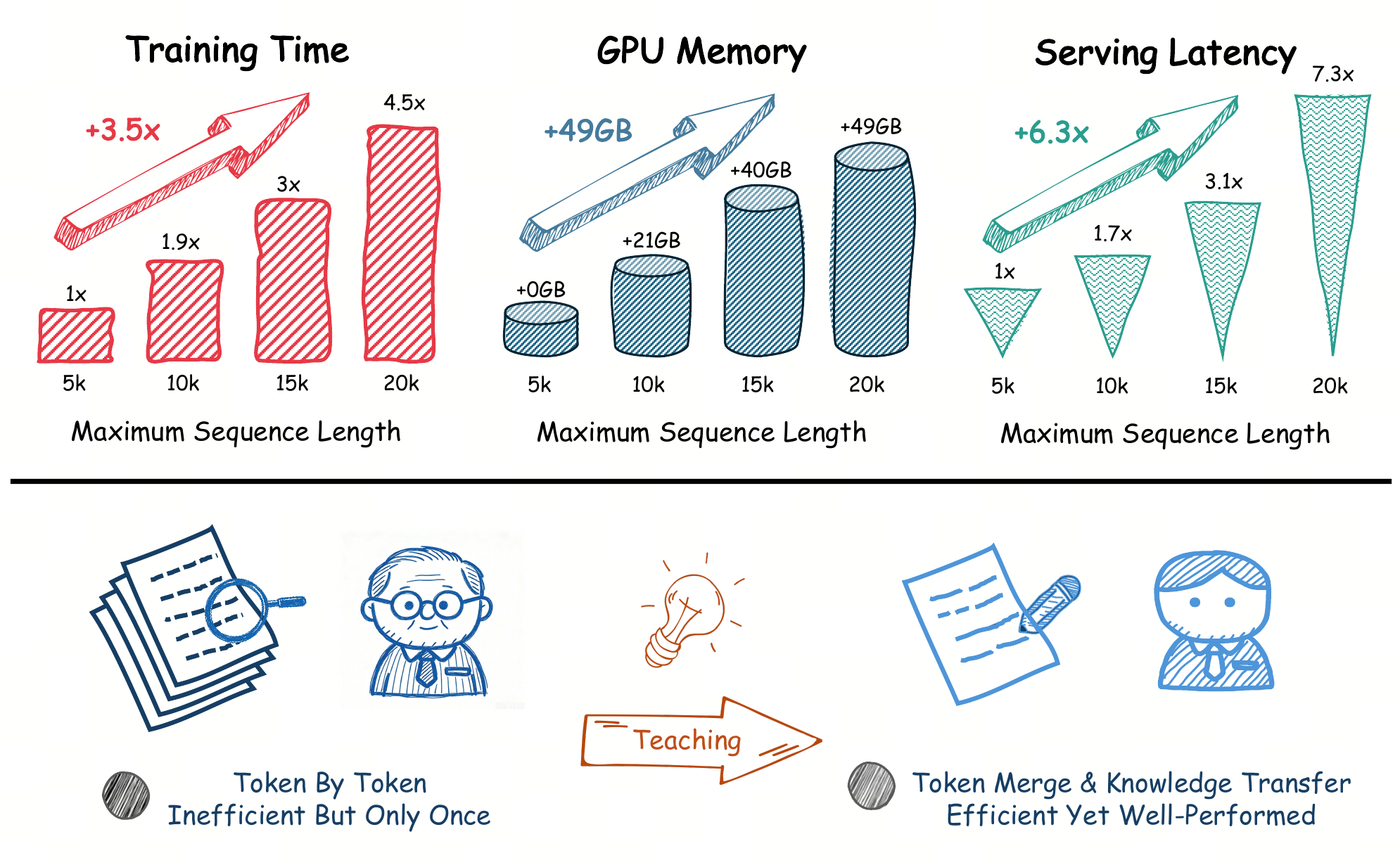}
  \caption{The motivation of TM20K. Ultra-long sequence modeling introduces significant burdens on training time, GPU memory, and serving latency. TM20K offloads heavy computation to a one-time trained teacher model that preserves all original tokens, enabling the student models to train and serve efficiently via well-motivated token merge strategies and the knowledge transferred from the teacher.}
  \label{fig:teaser}
\end{figure}

As one promising scaling direction, ultra-long user behavior sequence modeling has been widely shown to enhance recommendation quality in modern recommender systems and improve user satisfaction by simultaneously capturing users’ long-term and short-term interests~\cite{TWINV1,UniSRec,HiSAC,EST,IAT}. However, training and serving efficiency deteriorates dramatically as sequence length grows. For example, when extending the maximum sequence length from 5K to 20K in our real-world Ad recommendation system, even equipped with FlashAttention~\cite{FLA} and M-Falcon serving~\cite{GR,LONGER} for transformer modeling, training time increases by {\it $3.5\times$}, GPU memory increases {\it $49G$}, and serving latency increases by approximately {\it $6.3\times$} (shown in Fig.~\ref{fig:teaser}). That is, effectiveness and efficiency for long sequence modeling constitute conflicting objectives, demanding proper trade-offs to balance both aspects.

Existing recommender systems achieve this trade-off from various aspects, among which mainstream solutions fall into two categories: {\it sequence length compression} and {\it lightweight attention architecture}. Search-based mechanisms that retrieve a subset of highly relevant sequence tokens for each target candidate are widely adopted in conventional recommender systems~\cite{SIM,TWINV1,TWINV2,TransActV2,EST}, which inevitably discard fine-grained and potentially valuable behavioral information. Additionally, the search-based approaches are incompatible with M-Falcon serving~\cite{GR,LONGER}. Cluster-based methods can also aggregate ultra-long sequence tokens into condensed representations with a shorter length~\cite{TWINV2,VQL,ENCODE,HiSAC,DMQN}, yet they typically rely on high-quality pre-trained token embeddings to support stable clustering. Purely focusing on target-sequence attention or using varieties of lightweight attention mechanisms reduces the quadratic complexity of transformer modeling~\cite{TWINV1,TWINV2,STCA,EST,MTFM,HyTRec,HiSAC}. Despite their efficiency benefits, such lightweight attention fails to fully capture intrinsic dependencies among behavior tokens, leading to insufficient sequence feature extraction. A similar claim is proposed in the recent work~\cite{UltraHSTU}. In our real-world scenario, we indeed find that full transformer modeling exhibits obvious performance gains when compared with lightweight target attention. In addition to the above end-to-end solutions, {\it two-stage ranking frameworks} are more flexible to balance the effectiveness and efficiency of various model scaling directions, majorly including the paradigm of knowledge distillation~\cite{GapDis,ExLF,SUAN,RecDistill} and the foundation model~\cite{PretrainUE,LFM,LLaTTE,MARM,IAT,LoopFM,SIF}.

In this work, we adopt the {\it full attention (FA)} mechanism for sufficient and effective long sequence modeling rather than pure target attention (TA) based on preliminary experimental studies and observations. To tackle the substantial efficiency degradation caused by FA, we develop solutions from two perspectives: {\it two-stage knowledge distillation} and {\it sequence length compression via token merge}. Imagine a scenario in human education, as illustrated in Fig.~\ref{fig:teaser}: teachers enrich their knowledge by thoroughly reading the literature in full detail; in contrast, under the teacher's guidance, students may rapidly scan texts to complete their studies efficiently. Inspired by this analogy, we offload the heavy computation of full transformer computation to the teacher model, which {\it is trained only once and requires no online inference}. That is, the overhead introduced by the teacher model is nearly negligible under large-scale recommendation scenarios. For student models, we carefully design some token merge strategies to significantly compress the sequence length while avoiding noticeable performance degradation. By thoroughly analyzing the attention score distributions from an online model, we propose several simple yet well-motivated token merge strategies for student models, including:
\begin{itemize}[leftmargin=*]
    \item \textbf{Local-wise ID-based Token Merge (LITM)}. A full-scene e-commerce sequence plays a fundamental role in our Ad recommendation, which uses user behaviors on the interested products~\footnote{All data undergoes hash processing to safeguard user privacy.}. In practice, users may perform multiple interactions with the same product within a short time window. We aggregate these corresponding sequence tokens by product ID. LITM yields concentrated information representations, filters out anomalous behavioral noise, and further reinforces user privacy protection.
    \item \textbf{Position-wise Adaptive Token Merge (PATM)}. Prior research has demonstrated that recent sequence tokens show greater importance than the earlier ones~\cite{TransActV2,HyTRec,DV365,STCA}. Our internal analysis further reveals that the latest $10\%$ of tokens account for roughly half of all attention weights. Accordingly, we propose an adaptive token merge scheme that applies different merge intensities to sequence tokens based on their position index. By contrast, the prior work~\cite{LONGER} performs token merge uniformly for the entire sequence, assigning equal importance to recent and historical tokens.
    \item \textbf{Layer-wise Pyramid Token Merge (LPTM)}. Inspired by the attention score analysis performed in LLM~\cite{LLM-Layer1,LLM-Layer2}, we also observe distinct attention patterns across different transformer layers. Bottom layers exhibit evenly distributed attention scores, whereas scores in upper layers become more concentrated. Motivated by this observation, we progressively reduce the full transformer computation by adopting increasingly aggressive token merge strategies across the transformer layers.
\end{itemize}

\begin{table}[tb]
\renewcommand{\arraystretch}{1.2}
\caption{Comparisons of sequence length compression (LenCom) and attention architectures (AttnArch) with prior works. We also list the maximum raw sequence length (MaxLen) and the effective computed sequence length (EffLen) after length compression. Details are in Sec.~\ref{sec:relate-work}.}
\begin{tabular}{lcccc}
\toprule
\textbf{Method} & \textbf{MaxLen} & \textbf{LenCom} & \textbf{EffLen} & \textbf{AttnArch} \\ 
\midrule
TWIN-V1~\cite{TWINV1} & $10^4$ & GSU  &  $10^2$ & ETA   \\
TWIN-V2~\cite{TWINV2} & $10^6$ & HC \& GSU & $10^2$ & ETA  \\
EST~\cite{EST} & $10^6$ & SIM   & $10^3$ & LCA \& CSA  \\
STCA~\cite{STCA} & $10^4$ & LE   & $2\times 10^3$ & STCA  \\
HyTRec~\cite{HyTRec} & $10^4$ & - & $10^4$ & HA  \\
HiSAC~\cite{HiSAC} & $10^4$ & HC & $2\times 10^2$ & MHA  \\
HyFormer~\cite{HyFormer} & $3\times 10^3$ & - & - & CA  \\
MTFM~\cite{MTFM} & - & - & - & GQA \& HTA  \\
LONGER~\cite{LONGER} & $2\times 10^3$ & TM & $10^3$ & CA \& SA  \\ \midrule
TM20K-S & $2\times 10^4$ & TM & $2 \times 10^3$ & FA  \\
TM20K-T & $2\times 10^4$ & -  & $2\times 10^4$ & FA  \\
\bottomrule
\end{tabular}
\label{tab:tradeoff}
\end{table}

With the three well-motivated token merge strategies, student models themselves incur only marginal training and inference overhead when extending the maximum e-commerce sequence length from 5K to 20K, while delivering substantial performance gains. Benefiting from knowledge distillation, student models can achieve nearly $85\%$ of the performance obtained by a complete 20K sequence modeling. Tab.~\ref{tab:tradeoff} summarizes existing trade-off solutions for ultra-long sequences, covering closely related prior works as well as our proposed TM20K-S (Student) and TM20K-T (Teacher).

In summary, our main contributions are fourfold: \textit{(a) A Novel Two-Stage Knowledge Distillation Framework for Ultra-Long Sequence Modeling}. The one-time trained teacher retains full sequence tokens, and the student could merge tokens efficiently for online serving. \textit{(b) Several Simple Yet Well-Motivated Token Merge Strategies}. The proposed three types of token merge are easy to implement and deliver competitive predictive performance. \textit{(c) Remarkable Performance Improvements}. Extensive offline experiments validate the advantages of the proposed TM20K, and our real-world industrial recommender system achieves significant online metric gains with negligible additional overhead.

\section{Related Works}
\label{sec:relate-work}
\subsection{Sequence Length Compression}
Early works adopt hard-search or soft-search strategies to select candidate-relevant sequence items~\cite{SIM,DIN}. TWIN-V1~\cite{TWINV1} presents a consistency-preserving general search unit (CP-GSU) to enhance search quality. TWIN-V2~\cite{TWINV2} further leverages hierarchical clustering (HC) for users’ lifelong sequence modeling. Though these methods support a maximum sequence length of up to $10^6$, their effective computational lengths after compression remain relatively small, as summarized in Tab.~\ref{tab:tradeoff}. STCA~\cite{STCA} follows the length extrapolation (LE) paradigm by randomly dropping partial sequence tokens during training. HiSAC~\cite{HiSAC} also utilizes hierarchical clustering (HC). LONGER~\cite{LONGER} proposes a uniform token merge strategy for the entire sequence, ignoring the varying informativeness of sequence tokens. Instead, we {\it propose several well-motivated token merge schemes via carefully considering the tokens' properties}.

\subsection{Lightweight Attention Architecture}
TWIN~\cite{TWINV1,TWINV2} adopts efficient target attention (ETA), while STCA~\cite{STCA} leverages stacked target-to-history cross attention (STCA) for efficient sequence modeling. EST~\cite{EST} decomposes standard attention into lightweight cross attention (LCA) and content sparse attention (CSA). HyTRec~\cite{HyTRec} designs a hybrid attention architecture (HA) that integrates linear attention for long-term tokens and softmax attention for recent user interactions. MTFM~\cite{MTFM} incorporates grouped-query attention (GQA) and a customized hybrid target attention mechanism to lower computational complexity. LONGER~\cite{LONGER} further combines cross attention (CA) in bottom layers and self attention (SA) in upper layers following the Perceiver architecture~\cite{Perceiver}. Notably, {\it we take full attention for better sequence scaling} in this paper.

\subsection{Two-Stage Frameworks}
ExFM~\cite{ExLF} introduces the external large foundation model, which constructs a teacher like a foundation model (FM) that serves multiple students as vertical models (VMs) to amortize the building cost. This work has inspired several follow-up studies that adopt knowledge distillation (KD) to boost the student model's performance~\cite{SUAN,RecDistill}. Differently from these existing methods, we leverage KD to address the ultra-long sequence modeling challenge, where the teacher preserves full sequence tokens and student models achieve efficient training and inference. Foundation models equipped with various user embedding or item embedding caching have also been proven to improve downstream task performance~\cite{LFM,MARM,LLaTTE}. Recent concurrent studies have consistently demonstrated an emerging scaling paradigm termed instance-as-token modeling~\cite{IAT,SIF,LoopFM}. In this work, we {\it primarily focus on designing an effective knowledge distillation framework for ultra-long sequences, and leave foundation model integration as a promising future direction}.

\begin{figure*}[tb]
    \centering
    \subfloat[CVR AUC]{\includegraphics[width=0.23\textwidth]{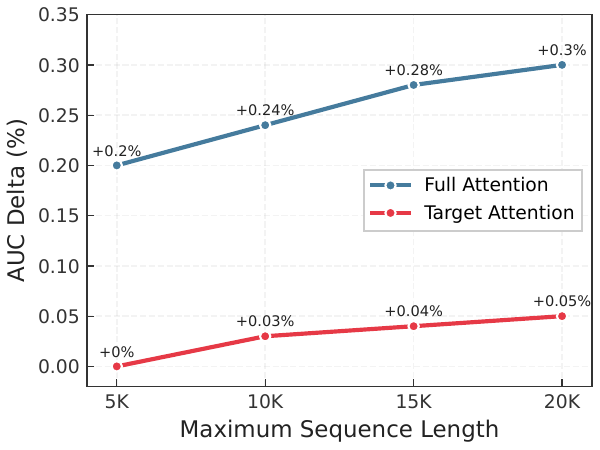}}
    \hspace{0.01\textwidth}
    \subfloat[Convergence (20K)]{\includegraphics[width=0.23\textwidth]{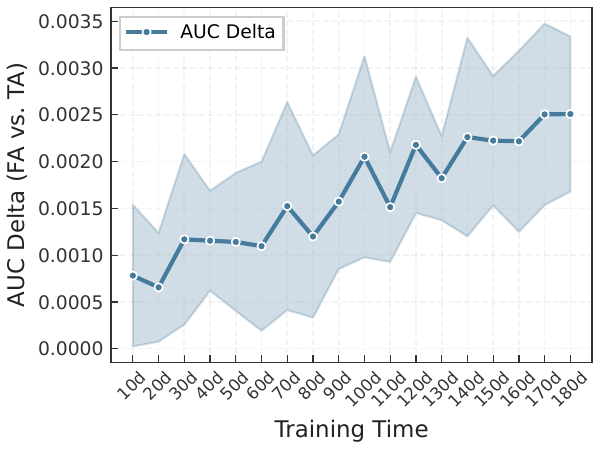}}
    \hspace{0.01\textwidth}
    \subfloat[GPU Utilization]{\includegraphics[width=0.23\textwidth]{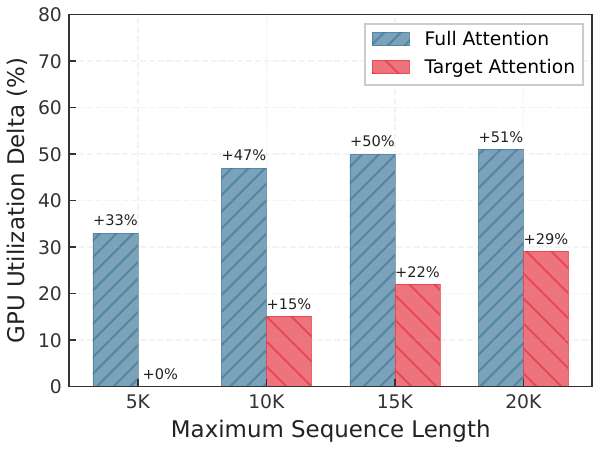}}
    \hspace{0.01\textwidth}
    \subfloat[Attention Distribution]{\includegraphics[width=0.23\textwidth]{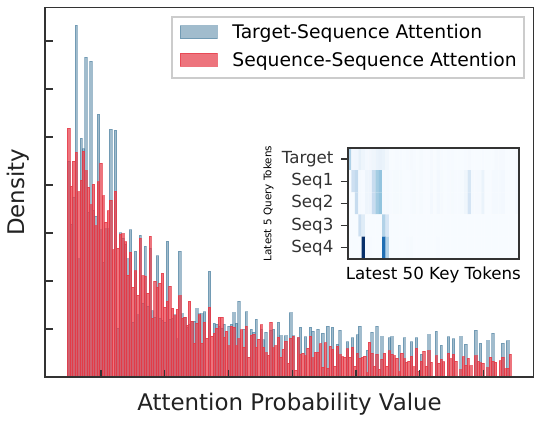}}
    \caption{Comparisons between full attention (FA) and target attention (TA) from multiple aspects.}
    \label{fig:fa_ta}
\end{figure*}

\section{Preliminaries and Observations}
We first introduce our real-world Ad recommendation system in Sec.~\ref{sec:problem}. We then compare full attention (FA) and target attention (TA) and explain our design choice in Sec.~\ref{sec:fa-ta}. Finally, Sec.~\ref{sec:attn-score} reports empirical observations on attention distributions that motivate our proposed methods.

\subsection{Problem Statement}
\label{sec:problem}
Existing industrial recommender systems commonly develop task-aware user behavior sequences to improve prediction performance~\cite{DIN,EST,TransActV2}. In our Ad recommendation scenario, users' historical interactions with products contribute significantly to conversion rate (CVR) prediction, which helps judge whether a user is interested in the target Ad. Such e-commerce sequence contains inherent product attributes (e.g., product ID, category), user-item interaction characteristics (e.g., interaction type), and contextual information (e.g., interaction timestamp). For simplicity, we use $S=\{s_1, s_2, \cdots, s_L\}$ to denote the sequence with $L$ tokens. Each token consists of a key ID feature $p_i$ (e.g., product ID) and other features $o_i$, formulated as $s_i=[p_i, o_i]$. For sequence modeling, we first embed each individual feature and aggregate token-level features via sum pooling:
\begin{equation}
    E_{s_i} = \text{SumPooling}\left(E_{p_i}, E_{o_i}\right),
\end{equation}
where $E_{p_i}$, $E_{o_i}$, and $E_{s_i}$ are $d$-dimensional vectors.
All token embeddings are further stacked to form the overall sequence tensor $E_{s}\in \mathbb{R}^{L \times d}$, which is fed into the sequence modeling module for target-aware feature extraction:
\begin{equation}
    x_{\text{seq}} = \mathcal{F}_{\text{seq}}\left(e_{t}, E_{s}\right),
\end{equation}
where $e_t \in \mathbb{R}^d$ represents the embedding of the target candidate item, and $x_{\text{seq}}$ encodes the interactive information between the target candidate and historical sequence tokens. After obtaining sequential representations, we fuse $x_{\text{seq}}$ with non-sequential features $x_{\text{non-seq}}$ and for subsequent feature interaction via RankMixer~\cite{RankMixer}:
\begin{equation}
    h = \mathcal{F}_{\text{interaction}}\left(x_{\text{non-seq}}, x_{\text{seq}}\right), \label{eq:rep}
\end{equation}
where the final representation $h$ is used to predict the conversion rate, and the model is updated through cross entropy loss.

\subsection{Full Attention vs. Target Attention}
\label{sec:fa-ta}
We conduct preliminary studies to determine the optimal sequence modeling architecture $\mathcal{F}_{\text{seq}}$. Most mainstream works rely on attention mechanisms~\cite{Transformer} but differ in their concrete implementations, as elaborated in Sec.~\ref{sec:relate-work}. Full attention (FA) is implemented as follows:
\begin{eqnarray}
    E_{c} &=& \text{Concat}\left(e_t, E_s\right), \label{eq:fa-concat} \\
    A_{c} &=& \text{Softmax}\left( \frac{E_cW_Q \left(E_cW_K\right)^T}{\sqrt{d}} + \mathcal{M}\right)E_cW_V, \label{eq:fa-softmax} \\
    O_{c} &=& \left(A_{c}W_u \odot \text{Swish}\left( A_c W_v\right)  \right)W_d, \label{eq:fa-ffn}
\end{eqnarray}
where $\mathcal{M}$ denotes the causal mask that forbids early tokens from attending to subsequent ones, which is also required by the M-Falcon serving~\cite{GR}. $W_{\star}$ denotes the learnable parameters of the transformer block. For simplicity, we omit layer normalization, multi-head attention and residual connection in the above equations.

In contrast, target attention (TA) simplifies the attention computation as follows:
\begin{eqnarray}
    a_{t} &=& \text{Softmax}\left( \frac{e_tW_Q \left(E_sW_K\right)^T}{\sqrt{d}}\right)E_sW_V, \label{eq:ta-softmax} \\
    o_{t} &=& \left(a_{t}W_u \odot \text{Swish}\left( a_t W_v\right) \right) W_d,
\end{eqnarray}
where only target-sequence attention is calculated, and only the target features are transformed by the SwiGLU FFN. Such inherent design leads to a drawback of {\it insufficient sequence feature extraction, especially for ultra-long sequences}.

We quantitatively verify this limitation via extensive empirical studies, as illustrated in Fig.~\ref{fig:fa_ta}. We build a 6-layer transformer backbone equipped with either FA or TA as the sequence modeling module, and conduct comparisons under various maximum sequence lengths (5K, 10K, 15K, and 20K). Fig.~\ref{fig:fa_ta}(a) reports the relative AUC gains over the baseline of 5K with TA. Fig.~\ref{fig:fa_ta}(b) further presents the improvement curves of FA over TA under the 20K sequence setting with over half a year of training data. The results demonstrate that FA consistently outperforms TA, achieving an AUC improvement of up to $0.25\%$ for the 20K sequence setting. It is worth noting that {\it the gain of FA on 5K sequences is even 0.15\% higher than that of TA on 20K sequences}, implying that the scaling benefit obtained through TA is not particularly substantial and that the performance gains from FA are by no means only attributable to increased computation. Beyond predictive performance, FA yields higher GPU utilization than TA as shown in Fig.~\ref{fig:fa_ta}(c). We further visualize the first-layer attention distributions of FA derived from Eq.~\ref{eq:fa-softmax} in Fig.~\ref{fig:fa_ta}(d). The target-sequence attention and sequence-sequence attention distributions are plotted respectively, which shows no obvious difference. A zoomed-in view of a local region containing 5 query tokens (including one target token and four sequence tokens) and 50 key tokens reveals that seq-seq interactions captures by FA carry essential predictive information that cannot be directly discarded. This phenomenon differs slightly from the findings of prior work~\cite{EST}. We speculate that search-based sequence compression adopted in their methods strengthens target-sequence interactions, thereby making sequence-sequence dependencies relatively less important. Through the above analysis, we {\it employ FA mechanism to support effective ultra-long sequence scaling, and pursue computational efficiency via alternative optimization perspectives}.

\begin{figure}[tb]
    \centering
    \subfloat[PID]{\includegraphics[width=0.145\textwidth]{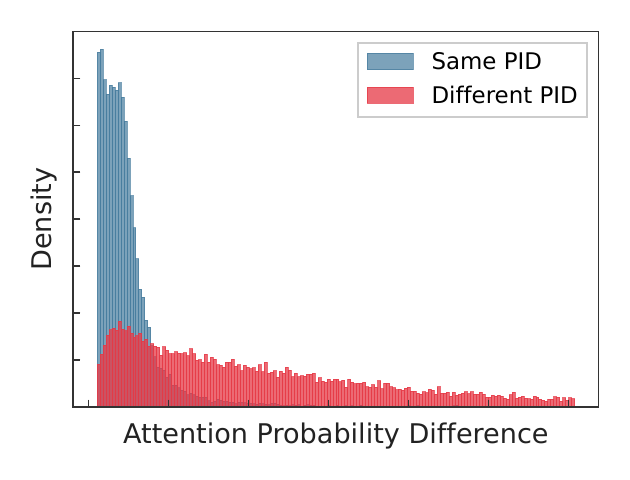}}
    \hspace{0.01\textwidth}
    \subfloat[Position Index]{\includegraphics[width=0.145\textwidth]{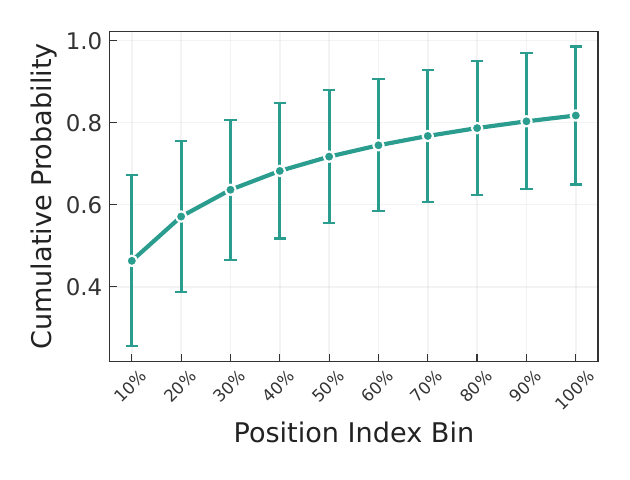}}
    \hspace{0.01\textwidth}
    \subfloat[Layer]{\includegraphics[width=0.145\textwidth]{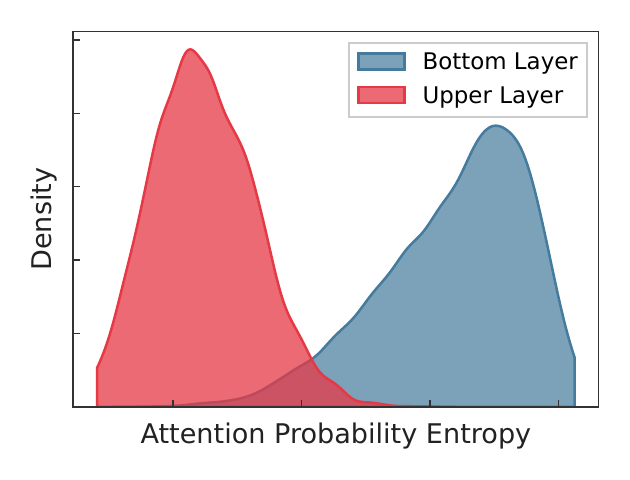}}
    \caption{Statistics and visualization of attention scores calculated from a subset of training instances.}
    \label{fig:attn_batch}
\end{figure}

\subsection{Attention Score Analysis}
\label{sec:attn-score}
We further conduct an in-depth analysis on the attention score distributions of a subset of training instances under the FA mechanism. The FA is trained with complete 20K sequence tokens, and we record a local submatrix of the attention probability matrix obtained via Softmax (Eq.~\ref{eq:fa-softmax}). For simplicity, we only report results from the first attention head. Let the attention submatrix be $A \in \mathbb{R}^{m \times n}$, where $m$ and $n$ denote the number of query tokens and key tokens, respectively. We set $m=5$ and $n=L=20,000$ to reduce the storage burden. The $n$ sequence tokens have product IDs denoted as $\{p_1,p_2,\cdots,p_L\}$.

First, we compute the absolute attention score difference for the $t$-th query token between paired key tokens, i.e., $|A_{t,i}-A_{t,j}|$. We plot the differences in Fig.~\ref{fig:attn_batch}, which is categorized into two groups based on whether $p_i$ equals $p_j$. Sequence tokens with same product ID exhibit closer attention scores. This {\it inspires us that the sequence tokens with same ID may be merged together}. Second, we partition the sequence tokens by their position indices, where a smaller position index bin corresponds to more recent interaction tokens. For each bin $[i_{k}, i_{k+1}]$, we aggregate the total attention weight via $\sum_{j=i_k}^{i_{k+1}}A_{t,j}$. The cumulative attention mass per bin is plotted in Fig.~\ref{fig:attn_batch}(b). The most recent top $10\%$ of sequence tokens contribute nearly half of the total attention weights. This finding motivates us to develop {\it an adaptive token merge strategy based on token positions}. Third, we measure the entropy of attention probability distribution (i.e., $-\sum_i A_{t,i}\log A_{t,i}$) across different layers. Fig.~\ref{fig:attn_batch}(c) shows that the upper layers yield sharper attention allocations over sequence tokens, whereas bottom layers tend to produce more uniform attention across sequence tokens. We accordingly try to {\it propose layer-wise progressive token merge schemes to reduce the computational overhead}. 


\begin{figure*}[tb]
  \centering
  \includegraphics[width=\linewidth]{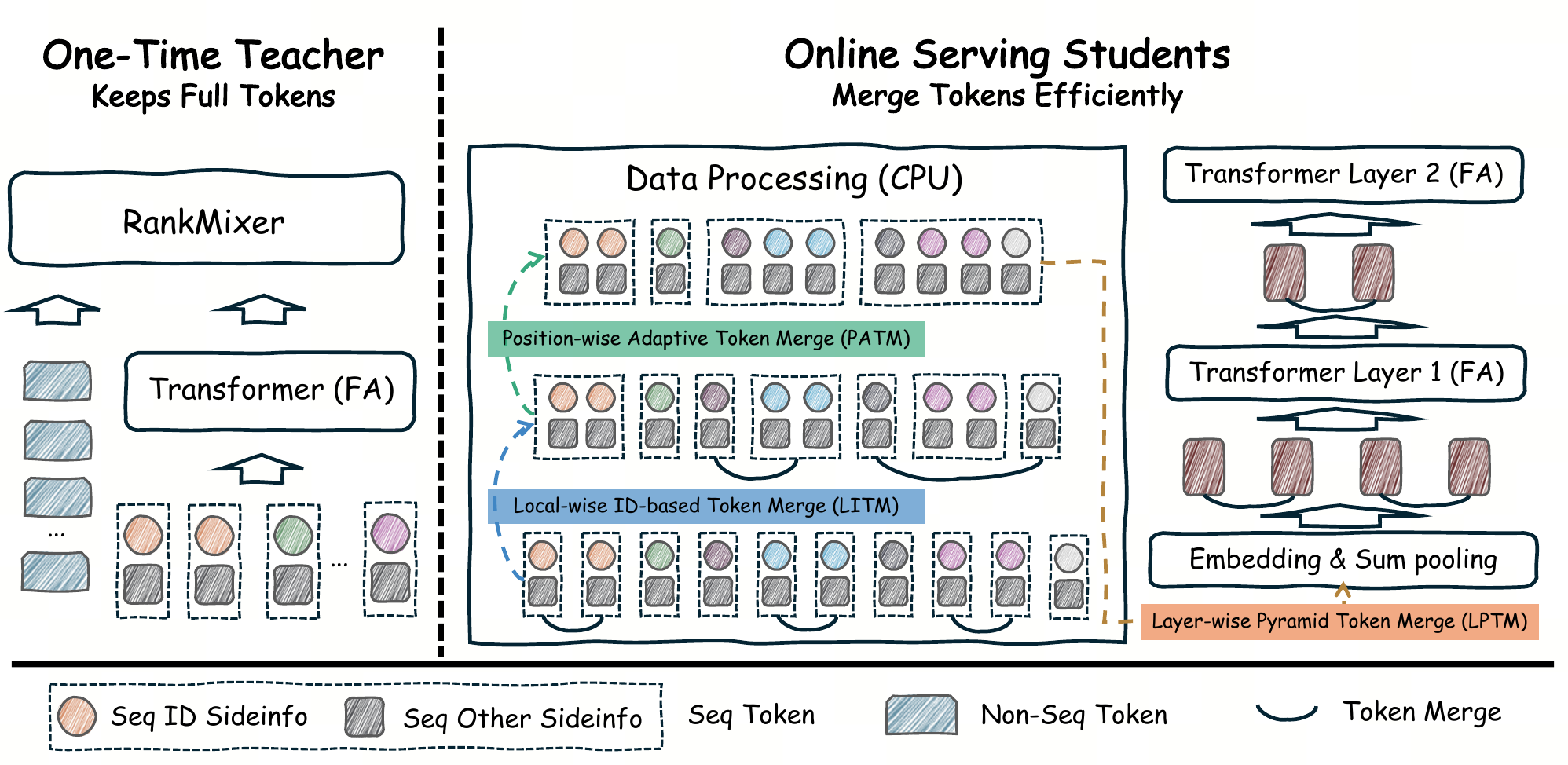}
  \caption{The overall framework of TM20K. Our model utilizes full attention (FA) for sequence modeling and RankMixer for heterogeneous token interaction. The teacher learns complete tokens without any token merge strategies, while the student models takes three simple yet well-motivated token merge strategies, including: (1) Local-wise ID-based Token Merge (\textbf{LITM}), (2) Position-wise Adaptive Token Merge (\textbf{PATM}), and (3) Layer-wise Pyramid Token Merge (\textbf{LPTM}).}
  \label{fig:method}
\end{figure*}

\section{Methodology}
This section elaborates our proposed tradeoff solution for 20K ultra-long sequence modeling within our industrial Ad recommendation system. We first overview the complete pipeline of TM20K in Sec.~\ref{sec:tm20k-framwork}. Next, we detail three simple yet principled token merge strategies tailored for student models in Sec.~\ref{sec:tm20k-s}. Third, we describe the one-time teacher model and present the KD workflow in Sec.~\ref{sec:tm20k-t}. Lastly, some additional designs for efficiency or stable training are provided in Sec.~\ref{sec:other}.
\subsection{TM20K Framework}
\label{sec:tm20k-framwork}
The overall framework of our proposed TM20K is illustrated in Fig.~\ref{fig:method}. We follow the KD framework~\cite{ExLF,SUAN,RecDistill} to enable more flexible ultra-long sequence scaling. Both teacher and student models adopt transformer modeling with full attention (FA) to capture comprehensive sequential information. As introduced in Sec.~\ref{sec:problem}, our model architecture utilizes RankMixer~\cite{RankMixer} for high-order interactions between sequential and non-sequential representations. {\it The teacher model keeps full sequence tokens without any sequence compression}. Although inefficiency, the teacher is trained only once and requires no online inference deployment. By contrast, {\it student models responsible for online serving process compressed sequences via three types of token merge approaches}, including: (1) Local-wise ID-based Token Merge (\textbf{LITM}), (2) Position-wise Adaptive Token Merge (\textbf{PATM}), and (3) Layer-wise Pyramid Token Merge (\textbf{LPTM}). 

\begin{algorithm}[htbp]
    \caption{Local-wise ID-based Token Merge (LITM)}
    \label{algo:litm}
    \begin{algorithmic}[1]
        \Require
            $E \in \mathbb{R}^{L \times d}$, $p \in \mathbb{I}^L$, $T$: maximum gap threshold
        \State Init $\texttt{p2i} \gets \emptyset$, $\texttt{G} \gets \emptyset$ \Comment{id2index map; temporal merged list}
        \For{$j = 0$ to $L - 1$} 
            \If{$p_j \notin \texttt{p2i}$ \textbf{or} $j - \texttt{p2i}[p_j] > T$}
                \State $\texttt{G}$.append($E_j$) \Comment{add new token}
                \State $\texttt{p2i}[p_j] \gets |\texttt{G}| - 1$ \Comment{record index}
            \Else
                \State $\texttt{G}[\texttt{p2i}[p_j]] \gets \text{Sum}(\texttt{G}[\texttt{p2i}[p_j]], E_j)$ \Comment{merge token}
            \EndIf
        \EndFor
        \State \Return $E_m \in \mathbb{R}^{L' \times d} \gets \texttt{G}$ \Comment{merged sequence}
    \end{algorithmic}
\end{algorithm}

\subsection{TM20K: Student Models}
\label{sec:tm20k-s}
Motivated by the empirical observations on attention scores (Sec.~\ref{sec:attn-score}), we devise three well-motivated token merge approaches for student models. The practical implementations are also provided.
\subsubsection{Local-wise ID-based Token Merge (LITM)}
Given the initial sequence token embeddings $E \in \mathbb{R}^{L \times d}$ and the key ID (e.g., product ID) list $p \in \mathbb{I}^{d}$. $\mathbb{I}$ is a discrete set that contains all unique IDs. A prior observation is that attention scores of sequence tokens with same ID tends to be more closer, which motivates us to merge these tokens together. Specifically, we propose local-wise ID-based token merge (LITM), which {\it continuously merge tokens with the same ID}. The pseudo-code is summarized in Algo.~\ref{algo:litm}. Here, $T$ serves as a hyperparameter that defines the maximum positional gap for merging tokens with identical key IDs. In our implementation, we fix $T=3$ to aggregate consecutive user interactions targeting the same key ID within a short local window.

\subsubsection{Position-wise Adaptive Token Merge (PATM)}
PATM is a position-wise token merge scheme, applying distinct compression strengths to tokens at different positions, guided by our prior attention analysis. It {\it preserves detailed information of recent tokens while aggressively compressing older and less informative ones}. We first divide the sequence length $L$ into $B$ disjoint segments defined by the range list $R$, where each entry $R_b$ is a tuple containing the start and end position indices. We apply the token merge operation proposed in~\cite{LONGER} to each segment with segment-specific compression factors stored in the vector of $K$. The pseudo-code is listed in Algo.~\ref{algo:patm}. Notably, we assume {\it tokens with smaller positional indices correspond to more recent user interactions}. Accordingly, the segment ranges and compression factors only need to satisfy the core principle: {\it recent behaviors undergo minimal token merge, while older historical interactions are compressed more aggressively}. For example, given $L=20,000$, $B=4$, segment ranges $R=[[1, 1,000],[1,001, 5,000],[5,001, 10,000],[10,001, 20,000]]$ and compression factors $K=[1, 2, 3, 4]$, the resulting merged sequence length is around 7,000. One implementation detail is that we pad the segment with all-zero tokens to ensure the length of it is divisible by its corresponding token merge factor.

\begin{algorithm}[htbp]
    \caption{Position-wise Adaptive Token Merge (PATM)}
    \label{algo:patm}
    \begin{algorithmic}[1]
        \Require
            $E \in \mathbb{R}^{L \times d}$,
            $B$,
            $R$,
            $K$
        \State Init $\texttt{G} \gets \emptyset$ \Comment{temporal merged list}
        \For{$b = 0$ to $B - 1$}
            \State $(s, e) \gets R_b$ \Comment{start and end position of segment $b$}
            \State $k \gets K_b$ \Comment{compression factor for current segment}
            \State $\hat{E} = \text{Sum}\big(\text{Reshape}\left(E[s:e], (-1, k, d)\right), \text{dim}=1\big)$ \Comment{merge $k$ consecutive tokens}
            \State $\texttt{G}$.extend($\hat{E}$)
        \EndFor
        \State \Return $E_m \in \mathbb{R}^{L' \times d} \gets \texttt{G}$ \Comment{merged sequence}
    \end{algorithmic}
\end{algorithm}

\subsubsection{Layer-wise Pyramid Token Merge (LPTM)}
LPTM reduces the computational overhead caused by FA through a progressive layer-wise token merge strategy. The sequence lengths across multiple layers forms a pyramid structure, where {\it bottom layers consume longer sequences while upper layers process much shorter sequences}. Suppose the $n$-th transformer layer's output is $\hat{E}_{n} \in \mathbb{R}^{L_n \times d}$, and then the $(n+1)$-th layer's input is merged as follows:
\begin{equation}
    E_{n+1} = \text{Sum}\left(\text{Reshape}\left( \hat{E}_n, \left(L_n/2, 2, d\right) \right), \text{dim}=1\right), \label{eq:lptm}
\end{equation}
where $E_{n+1} \in \mathbb{R}^{L_n/2, d}$ substantially reduces the computational cost of the subsequent attention layer. The pseudo-code is outlined in Algo.~\ref{algo:lptm}, where $N$ denotes the number of transformer layers. 
Only historical sequence tokens undergo layer-wise token merge, and the target candidate token is never merged. We omit this implementation in pseudo-code for brevity. Furthermore, the operation defined in Eq.~\ref{eq:lptm} {\it can be executed every multiple layers} to balance effectiveness and efficiency. In our scenario, we utilize a 6-layer transformer and performs this operation every 2 layers.

\begin{algorithm}[htbp]
    \caption{Layer-wise Pyramid Token Merge (LPTM)}
    \label{algo:lptm}
    \begin{algorithmic}[1]
        \Require
            $E_0 \in \mathbb{R}^{L_0 \times d}$,
            $N$
        \For{$n = 0$ to $N - 1$}
            \State $\hat{E}_n \gets \text{TransLayer}( E_n ) $ \Comment{according to Eq.~\ref{eq:fa-concat}-Eq.~\ref{eq:fa-ffn}}
            \State $E_{n+1} \gets \text{Merge}( \hat{E}_n ) $ \Comment{according to Eq.~\ref{eq:lptm}}
        \EndFor
        \State \Return $E_N \in \mathbb{R}^{L' \times d}$
    \end{algorithmic}
\end{algorithm}

\subsubsection{Practical Implementations}
We {\it in-order apply LITM, PATM and LPTM to compress 20K sequences} in our industrial Ad recommendation system, drastically reducing the computational cost of student models. We implement the computation of LITM and PATM on CPU. First, these rule-based token merge schemes are {\it CPU-friendly} and poorly suited for GPU acceleration. Furthermore, we avoid merging embeddings on the CPU, and instead, we {\it merge the raw features of each sequence token}. After grouping token features via LITM and PATM, we look up embeddings for each token's features from the embedding table and perform sum pooling afterward. This design significantly {\it reduces the overhead of communication bandwidth between CPU and GPU}. As for LPTM, it {\it only requires minor modifications to the standard transformer architecture}, and the merging operation formulated in Eq.~\ref{eq:lptm} can be efficiently implemented via tensor operations on GPU.

\subsection{TM20K: Teacher Model}
\label{sec:tm20k-t}
Because the teacher model is only trained once, we place little emphasis on its  efficiency. {\it Critically, the teacher model is trained independently of all student models, and its predicted logits are cached to supervise student training}. The teacher model follows the architecture introduced in Sec.~\ref{sec:problem}. Assume the predicted logit of the teacher is $g_{\text{T}} = \text{MLP}(h_{\text{T}})$, where $h_{\text{T}} \in \mathbb{R}^d$ denotes the final representation obtained in Eq.~\ref{eq:rep}. Then the teacher is optimized via the binary cross-entropy loss against the ground-truth label $y \in \{0, 1\}$:
\begin{equation}
    \ell_{\text{ce}} = -y \log q_{\text{T}} - (1-y)\log(1-q_{\text{T}}), \label{eq:teacher-ce}
\end{equation}
where $q_{\text{T}} = \text{Sigmoid}(q_\text{T})$ is the predicted conversion rate and is {\it cached for KD}.

Each student model is jointly optimized with the cross entropy loss and  distillation loss. Following the prior works~\cite{ExLF,RecDistill}, we design two distinct prediction heads for the student, including one main tower and one distillation tower formalized below:
\begin{equation}
    \ell_{\text{main}} = -y \log q_{\text{S,main}} - (1-y)\log(1-q_{\text{S,main}}), \label{eq:student-main}
\end{equation}
\begin{eqnarray}    
    \ell_{\text{dis}} &=& \ell_{\text{ce}} + \lambda \ell_{\text{kd}}, \label{eq:student-dis} \\
    \ell_{\text{ce}} &=& -y \log q_{\text{S,dis}} - (1-y)\log(1-q_{\text{S,dis}}), \\
    \ell_{\text{kd}} &=& -q_{\text{T}} \log q_{\text{S,dis}} - (1-q_{\text{T}})\log(1-q_{\text{S,dis}}).
\end{eqnarray}
Here, $q_{\text{S},\star}=\text{Sigmoid}\big(\text{MLP}_{\star}(h_{\text{S}})\big)$ with $\star \in \{\text{main}, \text{dis}\}$ representing the predicted conversion probabilities from the main tower and distillation tower, respectively. $\lambda$ is a hyperparameter.

\begin{table*}[htbp]
\centering
\renewcommand{\arraystretch}{1.2}
\caption{Overall performance and efficiency comparison across all methods. The maximum sequence length (MaxL), length compression strategy (LenCom), average and 90th percentile length of real computed sequences (AvgL and P90L) are reported. All AUC/LogLoss deltas are calculated relative to our 5K baseline model that has served online for a long time.}
\label{tab:overall_compare}
\resizebox{\textwidth}{!}{
\begin{tabular}{l c c c l c c c c c c}
\hline
\textbf{MaxL} & \textbf{LenCom} & \textbf{AvgL} & \textbf{P90L} & \textbf{Method} & \textbf{AUC($\uparrow$)} & \textbf{$\Delta$AUC(\%)} & \textbf{LogLoss($\downarrow$)} & \textbf{$\Delta$LogLoss(\%)} & \textbf{Thr} & \textbf{Mem} \\
\hline
\multirow{7}{*}{5K} & \multirow{7}{*}{\makecell{LITM0 \\ \& UTM2}} & \multirow{7}{*}{1.2K} & \multirow{7}{*}{1.6K} & Base & 0.8212 & -- & 0.4917 & -- & 88K & 52G \\
& & & & TWIN & 0.8159 & -0.65\% & 0.4984 & +1.37\% & 200K & 30G \\
& & & & STCA & 0.8200 & -0.15\% & 0.4931 & +0.29\% & 137K & 40G \\
& & & & HiSAC & 0.8207 & -0.06\% & 0.4922 & +0.11\% & 114K & 47G \\
& & & & LONGER & 0.8215 & +0.04\% & 0.4914 & -0.07\% & 100K & 58G \\
& & & & MTFM & 0.8213 & +0.01\% & 0.4916 & -0.02\% & 105K & 37G \\
& & & & HyFormer & 0.8216 & +0.05\% & 0.4913 & -0.08\% & 80K & 55G \\
\hline
\multirow{6}{*}{20K} & \multirow{6}{*}{UTM2} & \multirow{6}{*}{4.4K} & \multirow{6}{*}{10K} & TWIN & 0.8165 & -0.57\% & 0.4967 & +1.02\% & 102K & 53G \\
& & & & STCA & 0.8210 & -0.02\% & 0.4920 & +0.05\% & 58K & 68G \\
& & & & HiSAC & 0.8214 & +0.03\% & 0.4915 & -0.05\% & 15K & 63G \\
& & & & LONGER & 0.8221 & +0.11\% & 0.4907 & -0.21\% & 48K & 70G \\
& & & & MTFM & 0.8219 & +0.08\% & 0.4909 & -0.16\% & 27K & 81G \\
& & & & HyFormer & 0.8222 & +0.12\% & 0.4906 & -0.23\% & 36K & 75G \\
\hline
20K & TM20K & \textbf{8.8K} & \textbf{20K} & T & 0.8233 & \textbf{+0.26\%} & 0.4890 & \textbf{-0.55\%} & \textbf{11K} & 86G \\
\hline
20K & TM20K & \textbf{1.8K} & \textbf{2.6K} & S & 0.8224 & \textbf{+0.15\%} & 0.4903 & \textbf{-0.29\%} & \textbf{83K} & 74G \\
20K & TM20K & \textbf{1.8K} & \textbf{2.6K} & S w/ KD & 0.8230 & \textbf{+0.22\%} & 0.4896 & \textbf{-0.43\%} & \textbf{83K} & 74G \\
\hline
\end{tabular}
}
\end{table*}

\subsection{Other Designs}
\label{sec:other}
Our baseline model has leveraged a suite of optimization tools during training and inference, including User-Level Training (similar to RLB in~\cite{STCA}), FlashAttention~\cite{FLA}, Mixed-Precision Training~\cite{LONGER}, Remove-Padding, and M-Falcon Serving~\cite{GR}. We in addition introduce two optimizations:
\begin{itemize}[leftmargin=*]
    \item \textbf{Stack Sequence}. Sequences within a batch exhibit highly uneven lengths. Originally, all sequences are padded to match the longest sequence length when executing embedding table lookup, yielding an embedding tensor $E_s \in \mathbb{R}^{B \times L_{\text{max}} \times d}$. In practice, the total number of valid tokens $L_{\text{total}}=\sum_{i=1}^B L_{\text{valid},i}$ is often far smaller than $B \times L_{\text{max}}$, leading to substantial GPU memory waste. To mitigate this issue, we redistribute all valid tokens evenly across the batched instances and record the exact count of valid tokens for each instance. The obtained embedding tensor becomes $E_{\text{stack},s} \in \mathbb{R}^{B \times \overline{L} \times d}$ with $\overline{L}=\lceil L_{\text{total}}/B \rceil$. We then reconstruct the remove-padding tensor via the recorded token counts on the GPU afterward. This technique delivers prominent memory reduction for ultra-long sequences. For instance, {\it it reduces GPU memory usage by up to 10 GB for complete 20K sequences}.
    \item \textbf{QK Norm}. After incorporating ultra-long sequence modeling and KD loss, we observe unstable training in student models, which frequently suffers from training divergence. Drawing on solutions widely adopted in LLMs to mitigate instability induced by extreme attention scores~\cite{QKNorm}, we apply normalization layers to query and key tokens before attention computation. Although this introduces marginal overhead in GPU memory and computation, it {\it effectively stabilizes the entire training procedure}.
\end{itemize}

\section{Experiments}
This section first introduces the experimental settings, and then show the overall comparison results. Ablation studies about the proposed methods in TM20K are presented. Finally, we report the online A/B results in our real-world Ad recommendation system.

\subsection{Experimental Setup}
\label{sec:exper-setup}
\subsubsection{Datasets}
We conduct experiments on an industrial-scale CVR prediction dataset collected from a real-world advertising system. The training corpus spans two consecutive months and contains billions of training samples in total. The ultra-long user sequences consist of historical e-commerce interactions. Beyond sequence features, we feed user features, item features and contextual features into conventional feature interaction modules for further modeling. {\it All training data are anonymized by removing sensitive user/item information and hashing feature IDs}.

\subsubsection{Baselines}
The baseline model adopts a 6-layer transformer equipped with full attention (FA) to model user e-commerce historical sequences with a maximum input length of 5K. Our core objective is to {\it extend the supported maximum sequence length from 5K to 20K, aiming to achieve notable prediction performance gains without incurring excessive computational overhead}.  We compare our TM20K framework against the following state-of-the-art efficient sequence models. STCA~\cite{STCA} leverages stacked cross attention and length extrapolation to boost training efficiency. LONGER~\cite{LONGER} applies fixed uniform token merge (UTM) for sequence compression combined with the perceiver~\cite{Perceiver} architecture. MTFM~\cite{MTFM} adopts grouped-query attention and hybrid target attention to reduce sequence modeling complexity. HyFormer~\cite{HyFormer} designs a dedicated interaction block to model cross relationships between sequential and non-sequential features. We also run classic search-based and cluster-based sequential recommendation methods including TWIN~\cite{TWINV1} and HiSAC~\cite{HiSAC}.

\subsubsection{Hyperparameters and Evaluation Metrics}
The hidden dimension of transformer is 512. We adopt feed-forward networks (FFN) with SwiGLU activation, where the FFN intermediate dimension is set to 1024. The global training batch size is fixed at 320. For training the teacher model with full 20K sequence tokens, we use a smaller batch size of 96 due to the GPU memory burden. All experiments run on a distributed GPU cluster with hundreds of GPUs. The next-batch evaluation mechanism is used to calculate prediction metrics.
We evaluate all methods comprehensively by reporting both predictive performance and computational efficiency metrics. Area Under ROC Curve (AUC) and LogLoss serve as the prediction evaluation metrics. Training throughput (Thr) and peak GPU memory usage (Mem) quantify computational efficiency. We further report three statistical indicators of the processed input sequences under each training configuration, including maximum sequence length (MaxLen), average sequence length (AvgLen), and the 90th percentile sequence length (P90 Len). Due to the large volume of data used for model evaluation, we observe that the {\it AUC fluctuation across repeated model runs is within $0.01\%$}. Accordingly, standard deviations are omitted for all reported experimental results.

\subsection{Performance Comparisons}
\label{sec:exper-comp}
We conduct comprehensive comparisons covering the baseline sequence with MaxLen of 5K and extended sequence with MaxLen of 20K, and both performance and efficiency metrics are reported for all methods. The baseline sequence takes a simple version of ID-based token merge (denoted as LITM0) and uniform token merge with $K=2$ (denoted as UTM2)~\cite{LONGER} for processing 5K sequences. The results are listed in Tab.~\ref{tab:overall_compare}. TWIN~\cite{TWINV1} yields obvious performance drops due to its extremely simple sequence modeling paradigms (e.g., one interaction layer). 

On the 5K-length setting, STCA delivers inferior prediction performance with an AUC drop of 0.15\%. Under identical transformer depth and hidden dimension configurations, stacked cross attention cannot match the modeling capacity of vanilla full attention. The three other competitive long-sequence methods achieve mild accuracy gains of +0.04\%, +0.01\% and +0.05\%, respectively. In terms of efficiency, STCA and MTFM reduce GPU memory usage and increase training throughput, while HyFormer brings marginal accuracy gains at the cost of lower throughput and higher memory overhead. When extending the maximum sequence length to 20K, we adopt UTM2 compression for all comparison methods, as training with raw full-length 20K sequences would incur the explode of GPU memory. HiSAC~\cite{HiSAC} relies on well-trained multi-modal embeddings to support the clustering workflows, leading to small performance gains. Aside from STCA, other three compared approaches achieve positive AUC gains ranging from 0.08\% to 0.12\%, proving that longer user behavior sequences carry extra predictive signals. Nevertheless, efficiency deteriorates drastically, and all models suffer throughput declines of over 50\% compared to the 5K baseline model, limiting their practical online deployment.

Within our proposed TM20K framework, the teacher model TM20K-T captures the largest AUC improvement of +0.26\% and the lowest LogLoss (-0.55\%), confirming that complete uncompressed 20K sequences yield optimal prediction quality. However, its throughput collapses from the baseline 88K down to only 11K, with heavy GPU memory usage, making the teacher infeasible for online inference. Benefiting from the proposed three token merge approaches including LITM, PATM and LPTM, the student TM20K-S shrinks the average sequence length from the teacher’s 8.8K to merely 1.8K. Its standalone AUC gain decreases moderately to +0.15\%, yet throughput only falls slightly to 83K, corresponding to a throughput degradation of merely 5\% compared with the 5K baseline model. {\it Our token merge strategy drastically narrows the efficiency gap between long-sequence student and the original short-sequence baseline}. After introducing KD from the high-quality teacher logits, the student model TM20K-S with KD gains substantial predictive performance. Its AUC gain rises to +0.22\%, recovering around 85\% of the teacher’s total performance improvement. It is expected that further enhancing the teacher capacity would lead to additional gains for student models.

\begin{table}[htbp]
\centering
\renewcommand{\arraystretch}{1.2}
\caption{Ablation study on token merge strategies. Results are relative to the corresponding full-sequence baseline.}
\label{tab:ab_tm}
\resizebox{0.48\textwidth}{!}{
\begin{tabular}{l c c c c}
\hline
\textbf{TM Strategy} & \textbf{AvgL} & \textbf{P90L} & \textbf{$\Delta$AUC($\uparrow$)} & \textbf{Thr} \\
\hline
5K Full & 4K & 5K & -- & 40K \\
+UTM2 & 2K & 2.5K & -0.03\% & 72K \\
+LITM0 \& UTM2 & 1.2K & 1.6K & -0.05\% & 88K \\
\hline
20K Full & 8.8K & 20K & -- & 11K \\
+RandDrop & 5.3K & 13.7K & -0.11\% & 21K \\
+RandTrunc & 5.3K & 13.5K & -0.10\% & 21K \\
+LITM & 5K & 11.3K & -0.02\% & 25K \\
+PATM & 4.5K & 10.2K & -0.06\% & 34K \\
+LPTM & 8.8K & 20K & -0.03\% & 19K \\
+LITM \& PATM & 1.8K & 2.6K & -0.07\% & 68K \\
+LITM \& PATM \& LPTM & 1.8K & 2.6K & -0.11\% & 83K \\
\hline
\end{tabular}
}
\end{table}

\subsection{Ablation Studies}
\label{sec:exper-abla}
\subsubsection{Ablation on Token Merge Strategies}
\label{sec:exper-abla-tm}
We conduct ablation experiments to quantify the individual and combined impacts of LITM, PATM and LPTM on prediction performance and training efficiency, as summarized in Tab.~\ref{tab:ab_tm}. We first evaluate variants on the 5K full-sequence baseline by enabling UTM2 and LITM0 compression step by step. Applying UTM2 alone yields a mild AUC degradation of 0.03\% but improves training throughput from 40K to 72K. After stacking LITM0 with UTM2, the average sequence length is compressed to 1.2K. The overall AUC falls marginally by 0.05\%, while throughput is further lifted to 88K. We then shift to the 20K ultra-long sequence setting and benchmark each proposed token merge module against the uncompressed 20K full-sequence baseline. LITM shortens the average sequence length from 8.8K to 5K (a 40\% reduction), and PATM compresses it to 4.5K (a 48\% reduction). LPTM does not change the overall sequence length distribution but brings a small AUC drop of 0.03\% and noticeable throughput gains. Notably, we only report the sequence length of the first input layer, and the intermediate layers processes a shorter sequence with the help of LPTM. In terms of efficiency, LITM nearly doubles the throughput ($+1\times$), and PATM triples throughput ($+2\times$) relative to the 20K full-sequence baseline. Regarding predictive performance, standalone LITM, PATM, and LPTM lead to AUC declines of 0.02\%, 0.06\%, and 0.03\%, respectively. When all three strategies are integrated, the aggregated AUC loss reaches 0.11\%. That is, the introduced token merge strategies will inevitably cause conflicts in gradient updates, leading to performance degradation. However, they are well-supported by empirical observations from attention score distributions and have minimized information loss as much as possible. In contrast, if we randomly drop approximately 40\% of the sequence tokens using position-agnostic dropping (RandDrop) or random tail truncation (RandTrunc), the AUC loss already reaches around $-0.1\%$.


\begin{table}[htbp]
\centering
\caption{Hyperparameter analysis for LITM and PATM. Results are relative to the 20K full sequence.}
\label{tab:ab_hyper}
\resizebox{0.48\textwidth}{!}{
\begin{tabular}{l l c c}
\hline
 & \textbf{Hyperparameter} & \textbf{$\Delta$AvgLen($\downarrow$)} & \textbf{$\Delta$AUC($\uparrow$)} \\
\hline
\multirow{3}{*}{LITM} & $T=1$ & $-39\%$ & $-0.02\%$ \\
& \bm{$T=3$} & \bm{$-43\%$} & \bm{$-0.02\%$} \\
& $T=10$ & $-57\%$ & $-0.04\%$ \\
\hline
\multirow{3}{*}{PATM} & 2K:1,4K:2,8K:3,20K:4 & $-55\%$ & $-0.09\%$ \\
& \bf{3K:1,5K:2,10K:3,20K:4} & \bm{$-49\%$} & \bm{$-0.06\%$} \\
& 4K:1,8K:2,12K:3,20K:4 & $-40\%$ & $-0.05\%$ \\
\hline
\end{tabular}
}
\end{table}

\subsubsection{Ablation on Token Merge Hyper-parameters}
We conduct hyperparameter sensitivity experiments on LITM and PATM to explore the trade-off between sequence compression ratio and prediction performance, with detailed results listed in Tab.~\ref{tab:ab_hyper}. For LITM, we tune the gap threshold T among 1, 3 and 10. When T rises from 1 to 3, {\it the average sequence length is further compressed by an extra 4\% while the AUC remains the same}. For PATM, we test three groups of segment ranges and corresponding compression factors. For ease of notation, the configuration "2K:1,4K:2,8K:3,20K:4" indicates that we partition the full 20K sequence into four segments according to breakpoints of 2K, 4K, and 8K. Token merge is then performed on each segment with merging factors set to 1, 2, 3, and 4 respectively. A clear positive correlation is that more aggressive sequence compression consistently leads to larger AUC drop. For industrial online deployment, we select a moderately compressed PATM configuration. Our tuning principle {\it prioritizes controlling online serving latency within an acceptable range}, avoiding excessive efficiency deterioration even if slightly higher AUC loss could be tolerated.

\begin{table}[htbp]
\centering
\renewcommand{\arraystretch}{1.2}
\caption{AUC gains relative to the TM20K-S model with different distillation weights.}
\label{tab:ab_dis}
\resizebox{0.48\textwidth}{!}{
\begin{tabular}{l c c c c c}
\hline
& $\lambda$=30 & \bm{$\lambda=50$} & $\lambda$=75 & $\lambda$=100 & $\lambda$=150 \\
\hline
\textbf{$\Delta$AUC($\uparrow$)} & -0.01\% & \textbf{+0.00\%} & -0.01\% & -0.01\% & -0.03\% \\
\hline
\end{tabular}
}
\end{table}

\subsubsection{Ablation on Distillation Weights}
We investigate how different values of the $\lambda$ in Eq.~\ref{eq:student-dis} affect the student model’s AUC. The detailed results are summarized in Tab.~\ref{tab:ab_dis}. The optimal predictive performance is achieved when $\lambda$ is set to 50. Further inspection of loss magnitudes reveals that, under $\lambda=50$, the {\it scaled distillation loss term $\lambda\ell_{\text{kd}}$ has nearly identical magnitude to the loss $\ell_{\text{ce}}$}.

\begin{table}[htbp]
\centering
\renewcommand{\arraystretch}{1.2}
\caption{Results of other implementation details relative to the TM20K-S model.}
\label{tab:ab_other}
\resizebox{0.48\textwidth}{!}{
\begin{tabular}{l c c c}
\hline
& \textbf{Early $\Delta$AUC($\uparrow$)} & \textbf{Late $\Delta$AUC($\uparrow$)} & \textbf{Thr} \\
\hline
w/ MeanPooling \& LogN & -0.01\% & -0.01\% & +0\% \\
w/ PATM on SeqInc & +0.01\% & +0.01\% & +0\% \\
w/o QK Norm & +0.00\% & -0.45\% & +2\% \\
w/o KD in Late Period & +0.00\% & +0.00\% & +0\% \\
\hline
\end{tabular}
}
\end{table}

\subsubsection{Ablation on Other Implementations}
We conduct some auxiliary experiments and show results in Tab.~\ref{tab:ab_other}.
First, we replace sum pooling with mean pooling for token merge. Following prior work~\cite{TWINV2}, we incorporate $\log(n)$ into attention weights, where $n$ denotes the number of merged tokens. However, this modification causes an AUC degradation of 0.01\%. Second, we restrict PATM compression exclusively to the incremental tokens beyond the original 5K sequence when extending the length to 20K. This setup yields a marginal AUC improvement of 0.01\%, yet it substantially raises engineering complexity for model development. Third, we benchmark the performance and efficiency of training without QK normalization~\cite{QKNorm}. The early prediction performance remains roughly unchanged, while it may encounter training divergence as training proceeds. Lastly, we remove the distillation loss term at the late stage of training. The final prediction performance keeps nearly unchanged. This indicates that the student model has adequately absorbed knowledge from the teacher given sufficient distillation training iterations.

\begin{table}[htbp]
\renewcommand{\arraystretch}{1.2}
\caption{Online A/B results on our industrial Ad scene.}
\begin{tabular}{l c c c}
\toprule
 & \textbf{ADSS} & \textbf{ADVV} & \textbf{Latency} \\ 
\midrule
TM20K-S & +0.881\% (0\%)    & +0.515\% (0.004\%) &    +5.6\%   \\
TM20K-S w/ KD & +1.036\% (0\%) & +0.780\% (0\%)  &  +5.6\%   \\
\bottomrule
\end{tabular}
\label{tab:online_ab}
\end{table}

\subsection{Online A/B Results}
\label{sec:exper-ab}
Deploying the TM20K framework within our industrial advertising recommendation system achieves measurable improvements in core business metrics, namely Advertiser Score (ADSS) and Advertiser Value (ADVV). We allocate 10\% of the overall traffic for the online experiment, lasting five days and serving hundreds of millions of users. All online A/B test results are summarized in Tab.~\ref{tab:online_ab}. The baseline is a {\it mature high-performance advertising ranking model that has long been deployed online with the 5K e-commerce sequence modeling}. TM20K yields statistically significant positive gains across all business metrics, where {\it ADSS rises by 1.036\%, accompanied by a slight 5.6\% increase in online serving latency}. The p-values of the A/B tests are reported in parentheses within Tab.~\ref{tab:online_ab}, verifying that the observed gains are statistically significant. Notably, the teacher model has been deployed for up to {\it 8 months}, providing distillation signals for dozens of student model investigations, and {\it has not been retrained to date}. We think that retraining the teacher model once a year incurs a relatively small cost when amortized across all model training schedules. Furthermore, we utilize the memory-friendly M-Falcon~\cite{GR} strategy during online serving, where all target candidates share the same sequence information. Therefore, the serving cost primarily comes from the additional GPU cards required due to the increased latency (i.e., $5.6\%$), but {\it this is negligible compared to the online revenue gains}.

\section{Advantages and Limitations}
Our proposed TM20K framework has the following strengths: (a) the devised token merge strategies are {\it easy to develop and implement}, which are guided by solid observations derived from attention score analysis; (b) unlike prior methods relying on token retrieval or length extrapolation, TM20K {\it discards no tokens for either the teacher or student model}; (c) we validate that the two-stage distillation framework {\it enables ultra-long sequence scaling within an industrial large-scale recommendation system}, achieving substantial business improvements while maintaining favorable efficiency. 

Nevertheless, this work still has limitations. Specifically, the proposed token merge schemes are essentially {\it rule-based}. Although we conduct exhaustive ablation studies and hyperparameter analysis tailored to our advertising scenario, cross-domain deployment may still demand extra parameter tuning for adaptation to corresponding application scenarios. Furthermore, we have not yet incorporated {\it sparse attention mechanisms}~\cite{UltraHSTU,NSA,Kwai} into our scenes. Enabling compatibility between these sparse attention implementations and existing optimizations, including User-Level Training, Remove-Padding, and FlashAttention~\cite{FLA} {\it requires developing highly customized GPU operators}. This brings substantial engineering overhead and we leave the exploration of such sparse attention schemes for future work.

\section{Conclusion}
This paper proposes TM20K to obtain a balance between prediction performance and computational efficiency for ultra-long sequence modeling in an industrial advertising recommendation system. Comprehensive analysis of attention scores validate the necessity of adopting full attention (FA) for transformer modeling. To further mitigate the computational complexity introduced by FA, TM20K adopts a two-stage knowledge distillation paradigm. The teacher model is trained on complete sequence tokens, whereas the student leverages three well-motivated and carefully-designed token merge strategies to compress ultra-long sequences. By extending the maximum length of user e-commerce behavior sequences from 5K to 20K, TM20K delivers a statistically significant ADSS improvement of +1.036\% with negligible extra online serving latency.

\clearpage
\appendix




\bibliographystyle{ACM-Reference-Format}
\bibliography{tm20k}

\end{document}